\documentclass[5p,times]{elsarticle}

\usepackage[utf8]{inputenc}
\usepackage{amsmath}            % Para las referencias a ecuaciones con \eqref
\usepackage{epstopdf}           % Para poder insertar figuras .eps al compilar con PDFLATEX
\usepackage{flushend}           % Para igualar las columnas de la altima pagina
\usepackage{hyperref}           % Para hipervanculos dentro del PDF
\usepackage{amssymb}
\usepackage{lineno}

\biboptions{comma,round}
\usepackage{graphicx}  % needed for figures
\usepackage{dcolumn}   % needed for some tables
\usepackage{bm}        % for math
\usepackage{amssymb}
\usepackage{amsmath}   % for math
\usepackage{hyperref}
\usepackage[dvipsnames]{xcolor}
\usepackage{mathtools}
\hypersetup{colorlinks,linkcolor={blue},citecolor={blue},urlcolor={blue}}
\newcommand{\Mpl}{M_{_{\rm Pl}}}

\newcommand{\viz}{\textit{viz.~}}
\newcommand{\ie}{\textit{i.e.,~}}

\usepackage[figuresright]{rotating}

\usepackage{subfigure}
\begin{document}
	
	\begin{frontmatter}
		
		%% Title, authors and addresses
		
		%% use the tnoteref command within \title for footnotes;
		%% use the tnotetext command for the associated footnote;
		
		%% use the fnref command within \author or \address for footnotes;
		%% use the fntext command for the associated footnote;
		
		%% use the corref command within \author for corresponding author footnotes;
		%% use the cortext command for the associated footnote;
		%% use the ead command for the email address,
		%% and the form \ead[url] for the home page:
		%%
		%% \title{Title\tnoteref{label1}}
		%% \tnotetext[label1]{}
		%% \author{Name\corref{cor1}\fnref{label2}}
		%% \ead{email address}
		%% \ead[url]{home page}
		%% \fntext[label2]{}
		%% \cortext[cor1]{}
		%% \address{Address\fnref{label3}}
		%% \fntext[label3]{}
		
		%\dochead{Cabecera artaculo}
		%% Use \dochead if there is an article header, e.g. \dochead{Short communication}
		
		\title{Bounce from inflation}

		%% use optional labels to link authors explicitly to addresses:
		%% \author[label1,label2]{<author name>}
		%% \address[label1]{<address>}
		%% \address[label2]{<address>}
		
		\author{Debottam Nandi}
		\ead{debottam@physics.iitm.ac.in}
		%\ead[url]{www.cea-ifac.es}

		\address{Department of Physics, Indian Institute of Technology Madras, 
			Chennai 600036, India}
\begin{abstract}
We construct a class of viable bouncing models that are conformally related to cosmological inflation. There are three main difficulties in constructing such a model: (i) A stable (attractor) solution, (ii) A non-singular bounce, and (iii) to bypass the no-go theorem that states that simultaneously maintaining the observational bounds on the tensor-to-scalar ratio and the non-Gaussian scalar spectrum are not possible. We show that a non-minimal coupling of the scalar field helps to bypass these difficulties and provides a viable bouncing model with a naturally occurring reheating epoch briefly after the bouncing phase.
	
%In constructing a viable bouncing Universe, there exist three main difficulties: to construct (i) a stable (attractor) solution along with (ii) a non-singular bouncing phase, and (iii) to  circumvent the {\it no-go} theorem which states that simultaneously maintaining the observational bounds on the tensor-to-scalar ratio and the non-Gaussian scalar spectrum is not possible. In this letter, with the help of a simple non-minimal coupling, we construct a class of bouncing models which is conformal to the inflationary Universe. As a consequence, the above-mentioned difficulties are not present in our model. In addition to that, we obtain a naturally occurring reheating epoch briefly after the bouncing phase.
%This type of reconstruction implies that one may not be able to distinguish inflation and bouncing scenario from the observations as in different conformal frames, the Universe appears to be different.
%Conformal symmetry preserves the perturbations as well as the stability. Therefore, perturbed spectra in these bouncing models evade the {\it no-go} theorem and observational bounds remain the same as in the inflationary Universe, and also, the background solution becomes stable, leading to no BKL instabilities. In addition to those, there is a natural way of violating the null energy condition (which leads to the non-singular bouncing phase) and immediately after that, it enters into the conventional reheating phase.
\end{abstract}

\end{frontmatter}
%%%%%%%%%%%%%%%%%%%%%%%%%%%%%%%%%%%%%%%%%%%%%%%%%%%%%%%%%%%%%%%%%%%%%%%%%%%%%%%

\section{Introduction}

The inflationary paradigm remains the most successful \newline paradigm in explaining the early Universe as the theoretical predictions are in excellent agreement with the recent observations. However, there is an attempt of finding other alternatives to inflation as, despite the ever-tightening observational constraints, there seems to exist many inflationary models that continue remain consistent with the data \cite{Martin:2010hh, Martin:2013tda, Martin:2013nzq, Martin:2014rqa, Gubitosi:2015pba}. Also, it suffers from the trans-Planckian problem \cite{Martin:2000xs}. A popular alternative, free from the trans-Planckian problem is the classical bouncing scenario where, the Universe undergoes a phase of contraction until the scale factor reaches a minimum value, before it enters the expanding phase \cite{Novello:2008ra,  Cai:2014bea, Battefeld:2014uga, Lilley:2015ksa, Ijjas:2015hcc, Brandenberger:2016vhg}.

However, it also suffers difficulties, arguably even more than the inflationary paradigm. First, most of the bouncing solutions (except for ekpyrotic bounce) are not stable and therefore, in these cases, the anisotropic energy density can grow faster than the energy density responsible for the bouncing scale factor solution (contracting phase),  and hence may potentially break the homogeneous background, known as the  Belinsky-Khalatnikov-Lifshitz (BKL) instability \cite{doi:10.1080/00018737000101171, Cai:2013vm}. Second, while constructing the contracting phase is rather easy to achieve, obtaining the non-singular bouncing phase is extremely difficult as it requires to violate the null energy condition: often called as the \emph{theoretical} no-go theorem \cite{Kobayashi:2016xpl, Libanov:2016kfc, Ijjas:2016vtq, Banerjee:2018svi, Cai:2016thi, Cai:2017dyi, Kolevatov:2017voe, Mironov:2018oec, Easson:2011zy, Sawicki:2012pz}. Third, and most importantly, even if one may construct a model evading the first two difficulties, the perturbations suffer from many difficulties (e.g., gradient instability) and these models fail to be in line with the observational constraints: a small tensor-to-scalar ratio $(r_{0.002} \lesssim 0.06)$ and simultaneously, very small scalar non-Gaussianity parameter $\left(f_{\rm NL} \sim \mathcal{O} (1)\right)$: referred to as the \emph{observational} no-go theorem \cite{Quintin:2015rta, Li:2016xjb}. Apart from these main three issues, in general, bouncing models possess another, rather weak, difficulty as well: a natural exit mechanism from the bouncing phase to enter into the conventional reheating phase.

In solving these problems, it is soon realized that one needs to go beyond the canonical theories and consider non-minimal couplings, \viz the Horndeski theories or even beyond Horndeski theories \cite{Horndeski:1974wa, Gleyzes:2014dya, Kobayashi:2019hrl, Cai:2013vm, Kobayashi:2016xpl, Libanov:2016kfc, Ijjas:2016vtq, Banerjee:2018svi, Cai:2016thi, Cai:2017dyi, Kolevatov:2017voe, Mironov:2018oec, Kobayashi:2016xpl, Quintin:2015rta, Li:2016xjb, Cai:2012va, Ilyas:2020qja, Dobre:2017pnt}.
%However,  recently, in order to understand and deal with the problem regarding the stable solution, it has been proposed that the issue can be resolved by using the simplest non-minimal coupling \cite{Nandi:2018ooh, *Nandi:2019xlj} and at the same time, {\it consistency relation} (a feature that is violated in most bouncing models while satisfied in the inflationary scenario) can be restored in bounce \cite{Nandi:2019xag}. These results are obtained by using the conformal transformation.
However, it is still an open issue to construct a viable bouncing model that can evade all the above-mentioned difficulties, simultaneously. This is because: {\it there is not a single mechanism known that can be used to resolve all the issues at the same time.}

However, our recent works suggest that conformal coupling can play a pivotal role in resolving those issues \cite{Nandi:2018ooh, Nandi:2019xlj, Nandi:2019xag}. With that keeping in mind, in this letter, {\it we intend to construct a (class of) model that is stable and at the same time, may provide viable perturbed spectra with the help of conformal coupling.} The approach is the following: \emph{since most of the slow-roll inflationary models satisfy all the observational constraints, we shall conformally transform the inflationary scale factor solution into a bouncing solution.} Since the conformal transformation preserves the stability and the perturbations, the reconstructed bouncing model also holds the stability as well as perturbations, similar to the conformal inflationary model. Later, we shall also show that this kind of reconstruction helps us to automatically achieve a non-singular bouncing phase as well as a naturally occurring exit mechanism from bounce to reheating epoch, hinting towards a class of \emph{the first viable bouncing models.}

\section{Constructing the model} Slow-roll inflationary model can easily be constructed by using a single canonical scalar field $\phi$ minimally coupled to the gravity, i.e., the Einstein gravity as

\begin{eqnarray}\label{Eq:MinAc}
	\mathcal{S}_{I} = \frac{1}{2} \int {\rm d}^4{\rm \bf x} \sqrt{-g_{ I}} \left[\Mpl^2 R^I - g_{I}^{\mu \nu} \partial_\mu \phi \partial_\nu \phi - 2\,V_{I}(\phi)\right].\quad
\end{eqnarray}

\noindent The sub(super)script `$I$' (not to be confused with power or any indices) denotes the quantity in the minimal Einstein theory (similarly, we shall reserve `$b$' for the yet to be constructed bouncing theory), $R^I$ is the Ricci scalar for the metric $g^I_{\mu \nu}$, and $V_{I}(\phi)$ is the potential responsible for the inflationary solution. Assuming the above theory is responsible for the slow-roll inflationary dynamics, the scale factor solution, during inflation, can {\it approximately} be written as a function of the scalar field $\phi$ as
%$a_I(\eta)$ in comoving time $\eta$ can be approximated as $a_I(\eta) \approx - 1/\left(H_I\,\eta\right)$, where $H_I$ is the inflationary Hubble parameter (nearly constant). It also can {\it approximately} be written as a function of the scalar field $\phi$ as

\begin{eqnarray}\label{Eq:scaleInf}
	a_{I}(\phi) \propto\,\, \exp \left(-\int^\phi \frac{\mathrm d \phi}{\Mpl^2}\frac{V_I}{V_{I, \phi}}\right),
\end{eqnarray}

\noindent with $V_{I, \phi} \equiv \frac{\partial V_I}{\partial \phi}$. Now we shall construct a model which is conformal to the above inflationary action \eqref{Eq:MinAc} in such a way that the new scale factor behaves as a bouncing solution. The transformation can be written as
\begin{eqnarray}\label{Eq:ConfTGen}
g^{I}_{\mu \nu} = f^2(\phi)\,g^{b}_{\mu \nu} \quad \Rightarrow \quad a_I(\eta) = f(\phi)\,a_b(\eta).
\end{eqnarray}

\noindent %$g^I_{\mu \nu}$ and $a_I(\eta)$ are the inflationary metric and the scale factor solutions for the action \eqref{Eq:MinAc}, respectively, and 
$g^{b}_{\mu \nu}$ and $a_b(\eta)$ are the required bouncing metric and scale factor solutions, respectively, along with $f(\phi)$ being the coupling function. Using the above conformal transformation along with the action \eqref{Eq:MinAc}, we can construct the action responsible for such bouncing solution as
\begin{eqnarray}\label{Eq:NonMinAc}
\mathcal{S}_{b} &=& \frac{1}{2} \int {\rm d}^4{\rm \bf x} \sqrt{-g_{b}} \left[\Mpl^2\, f^2(\phi)\,R^b -\omega(\phi) \,g_{b}^{\mu \nu} \partial_\mu \phi \partial_\nu \phi \right.\nonumber\\
&&\left. \qquad \qquad\qquad\qquad\qquad\qquad - 2\, V_{b}(\phi)\right].
\end{eqnarray}

\noindent %$R^b$ is the new Ricci scalar for the bouncing metric $g^b_{\mu \nu}$ and the solution to the above action leads to the required bouncing scale factor solution $a_b(\eta)$.
$\omega(\phi)$ and the potential $V_b(\phi)$  depend on the coupling function $f(\phi)$ and the inflationary potential $V_I(\phi)$ as
\begin{eqnarray}\label{Eq:Potb}
\omega(\phi) = f^2(\phi)\left(1 -  6 \Mpl^2\, \frac{f_{,\phi}{}^2}{f^2}\right),~ V_b(\phi) = f^4(\phi)\,V_I(\phi),\quad\,
\end{eqnarray}
\noindent where $f_{, \phi} \equiv \frac{\partial f}{\partial \phi}$.

For a given inflationary model \eqref{Eq:MinAc} with its solution \eqref{Eq:scaleInf}, one can set $f(\phi)$ in such a way that $a_b(\eta)$ behaves as a bouncing solution.
%$f(\phi)$ is the conformal coupling function and depending upon the functional dependence of $f(\phi)$, one can easily obtain the scale factor solution $a_b(\eta)$.
For simplicity, if we desire the bouncing solution of the form
\begin{eqnarray}\label{Eq:scaleb}
a_b(\eta) \propto (-\eta)^\alpha \propto \exp\left(\alpha\,\int^\phi \frac{\mathrm d \phi}{\Mpl^2}\frac{V_I}{V_{I, \phi}}\right), \quad \alpha > 0,
\end{eqnarray}
\noindent where $\eta$ is the comoving time, then by using Eq. \eqref{Eq:scaleInf} and \eqref{Eq:ConfTGen}, one can obtain the required solution of the coupling function $f(\phi)$ as

\begin{eqnarray}\label{Eq:coupling}
f(\phi) = f_0\,\exp\left(-\frac{\left(\alpha + 1\right)}{\Mpl^2} \int^\phi \mathrm d \phi\,\frac{V_I}{V_{I, \phi}}\right).
\end{eqnarray} 
\noindent $f_0$ is normalized in such a way that at the minima of the potential, $f(\phi)$ becomes unity. {\it This is the main result of the work:} given an inflationary potential $V_I(\phi)$, we can completely determine the coupling function $f(\phi)$, which in turn, provides non-minimal theory \eqref{Eq:NonMinAc} responsible for the  bouncing scale factor solution \eqref{Eq:scaleb}. Note that, since Eq. \eqref{Eq:scaleInf} is an approximated form, by using the above expressions \eqref{Eq:coupling} and \eqref{Eq:ConfTGen}, the obtained scale factor solution $a_b(\eta)$ is also not exact, but close to \eqref{Eq:scaleb}. Also, we need to stress that the form of the scale factor \eqref{Eq:scaleb} is valid only in the contracting phase, and during and after the bounce, it is extremely difficult to obtain analytical solution. Therefore, the bouncing model, in our case, is also {\it asymmetrical}.

\par  Since conformal transformation preserves the stability (for details, see Refs. \cite{Nandi:2018ooh, Nandi:2019xlj}), the reconstructed bouncing model \eqref{Eq:NonMinAc} is as stable as the inflationary model and thus can evade BKL instability as well. In case of scalar perturbation, in the Einstein frame \eqref{Eq:MinAc}, the perturbed action is $\int {\rm d}\eta\, {\rm d \bf x}^3 z_I^2 \left(\zeta_I{}^\prime{}^2 - \left(\nabla \zeta_I\right)^2 \right),$ $z_I(\eta) \equiv  \frac{a_I \phi^\prime}{\mathcal{H}_I},$ where, $\zeta$ is the curvature perturbation. In the non-minimal frame \eqref{Eq:NonMinAc}, the expression of the action for the scalar perturbation changes by replacing $z_I(\eta)$ with $z_b(\eta)$. However, $z_b(\eta)$ is simply the conformally transformed $z_I(\eta)$, i.e., $z_b (\eta) = (f(\phi) \,a_b(\eta)\, \phi^\prime)/(\mathcal{H}_b + \phi^\prime f_{,\phi}(\phi)/f(\phi))$. This implies that, $z_b(\eta) = z_I(\eta)$ which leads to the action being identical in both frames and hence, $\zeta_I = \zeta_b$ at linear order. Also, as the speed of sound is conformally invariant, $c_s$ is unity in both minimal inflationary as well as in non-minimal bouncing scenarios. Similarly, one can evaluate the perturbed interaction Hamiltonian (for detailed evaluation, see Refs. \cite{Chen:2006nt, Maldacena:2002vr, Nandi:2015ogk, Nandi:2016pfr, Nandi:2017pfw}) at any order and show that, it also remains invariant in all conformal frames.  This implies that the curvature perturbation at any order in both the frames are identical, which is not surprising, as we know, {\it under conformal transformation, curvature perturbation remains invariant.} The argument extends to tensor perturbation as well. Therefore, given a viable inflationary model which is consistent with the observations, in our bouncing model, there appear no divergences or instabilities (e.g., gradient instability) in the perturbations anywhere, even at the bounce and, in addition to that, the model also satisfies all observational constraints: thus evading the no-go theorem.

%This is the main result of this work: \emph{given a slow-roll inflationary model \eqref{Eq:MinAc}, with the help of the chosen coupling function \eqref{Eq:coupling}, one can construct a class ($\alpha$ is arbitrary) of non-minimal theories \eqref{Eq:NonMinAc} that drives the required bouncing scale factor solutions \eqref{Eq:scaleb}. These class of models are as stable and the perturbations remain the same as in the case of inflationary counterpart. Also, we shall show that the non-singular bouncing phase as well as the reheating epoch appear automatically in our model.}

\begin{figure}[!t]
	\centering
	\includegraphics[width=.9\linewidth]{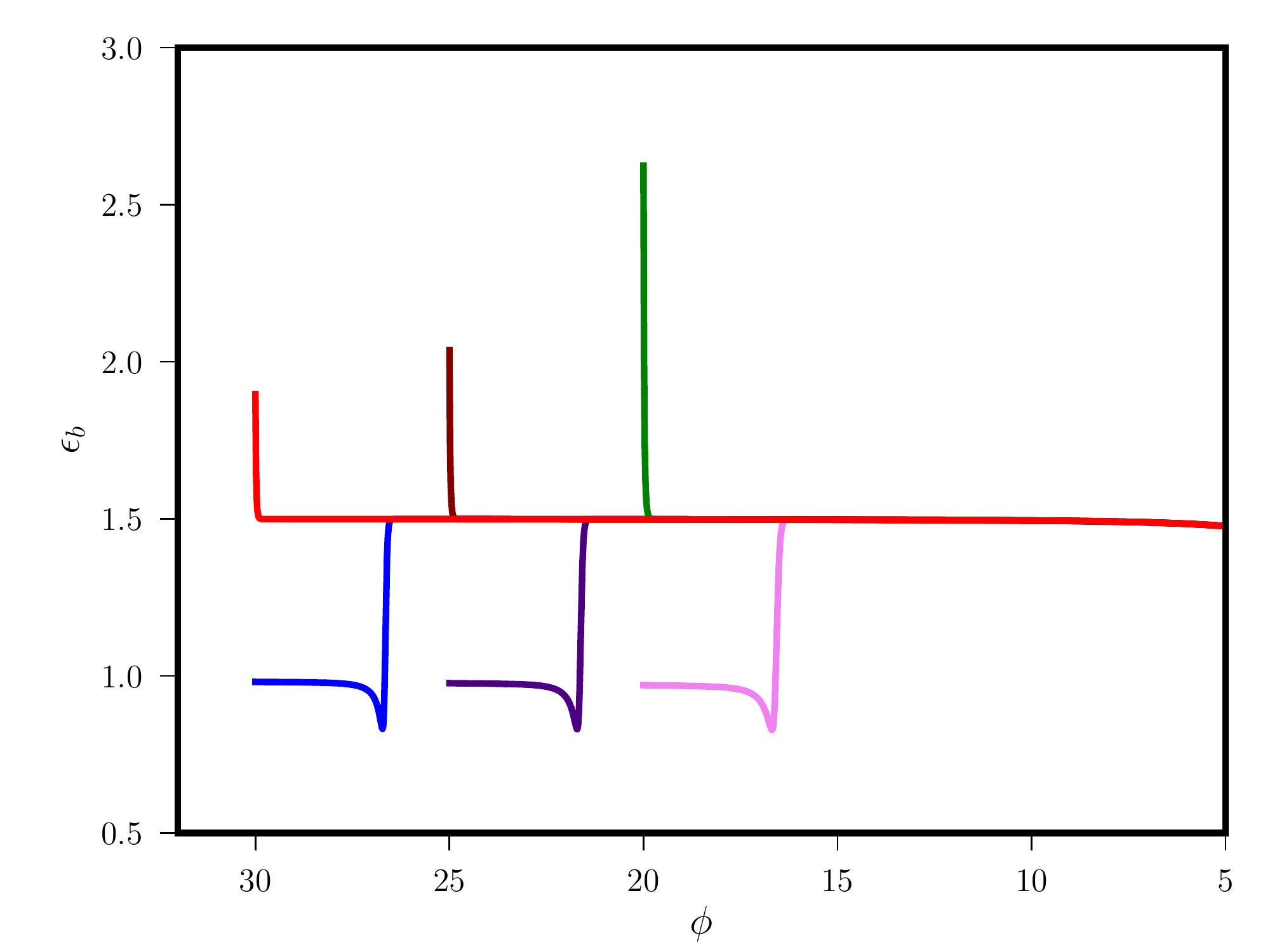}
	\caption{Slow-roll parameter $\epsilon_b$ is plotted for the chaotic bouncing model with $\alpha = 2$ with respect to the field $\phi$. Different colors signify different initial conditions (open-ended points) which is far away from the desired value of the slow-roll parameter. However, all solutions quickly approach to the value of $\epsilon_b \simeq 3/2$ which is the solution for $\alpha = 2$. This clearly indicates that our bouncing model is an attractor.}\label{Fig:eos1}
\end{figure}

\begin{figure*}[!t]
	\centering
	\includegraphics[width=.4\linewidth]{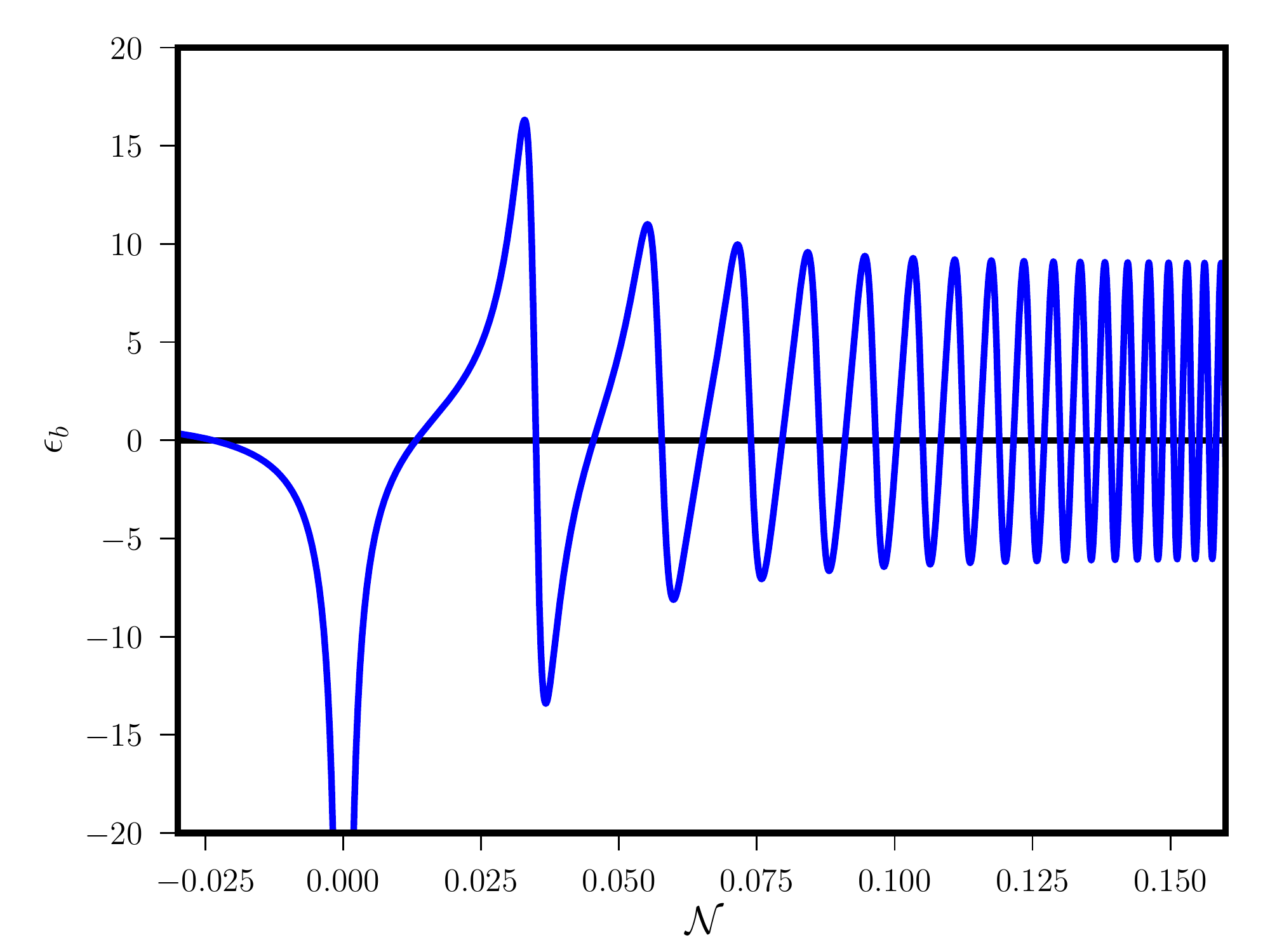}
	\includegraphics[width=.4\linewidth]{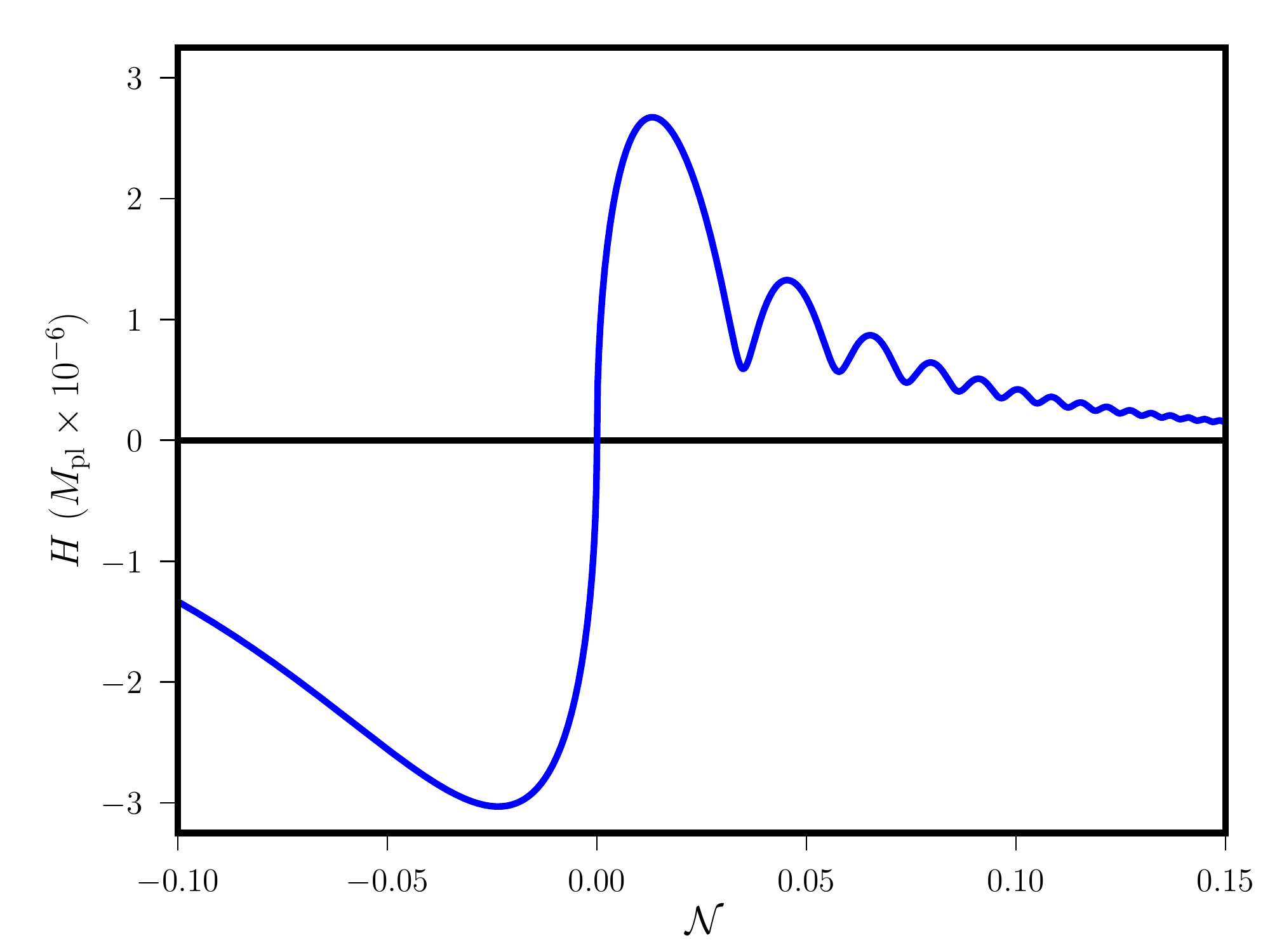}
	\includegraphics[width=.4\linewidth]{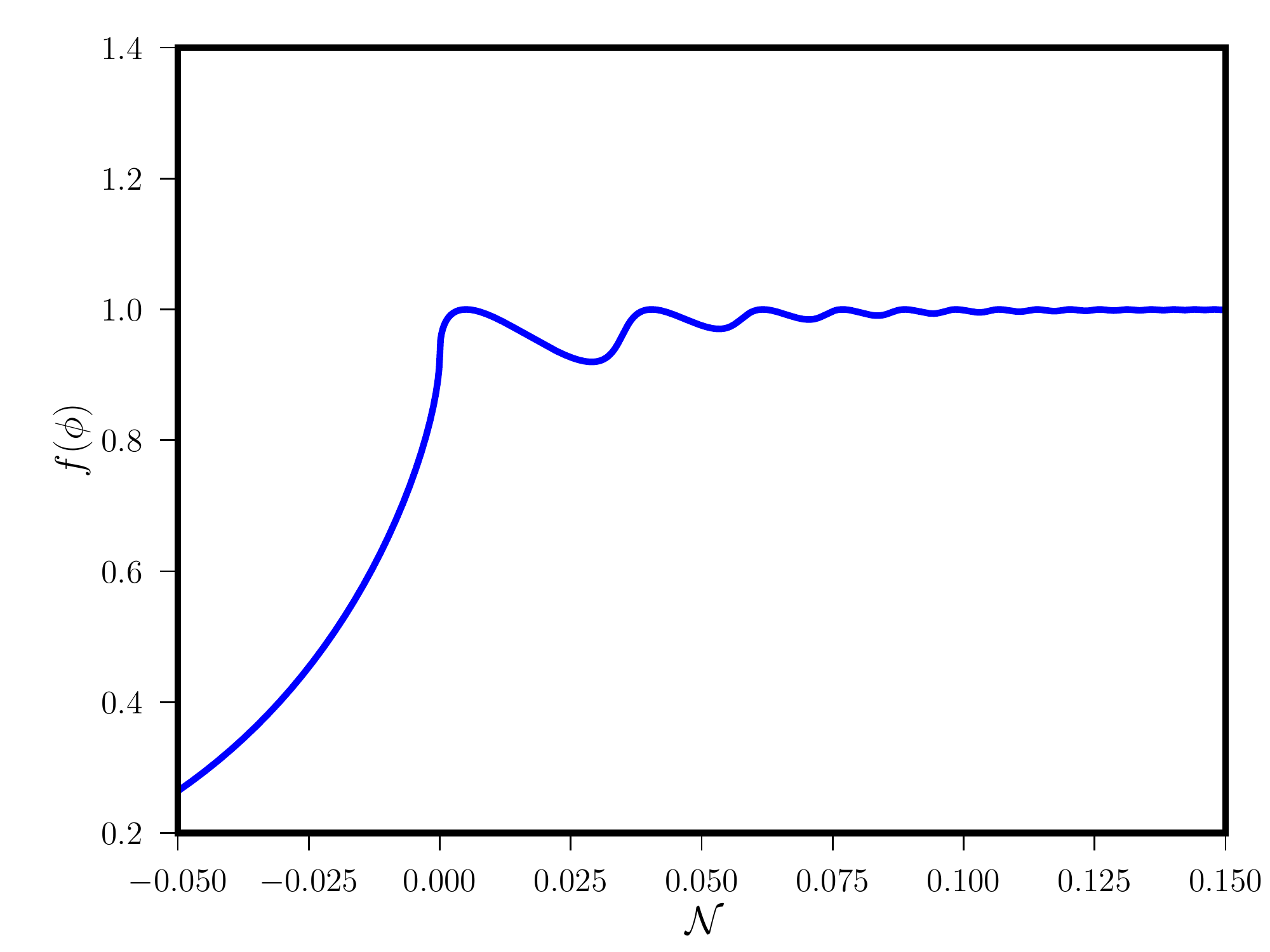}
	\includegraphics[width=.4\linewidth]{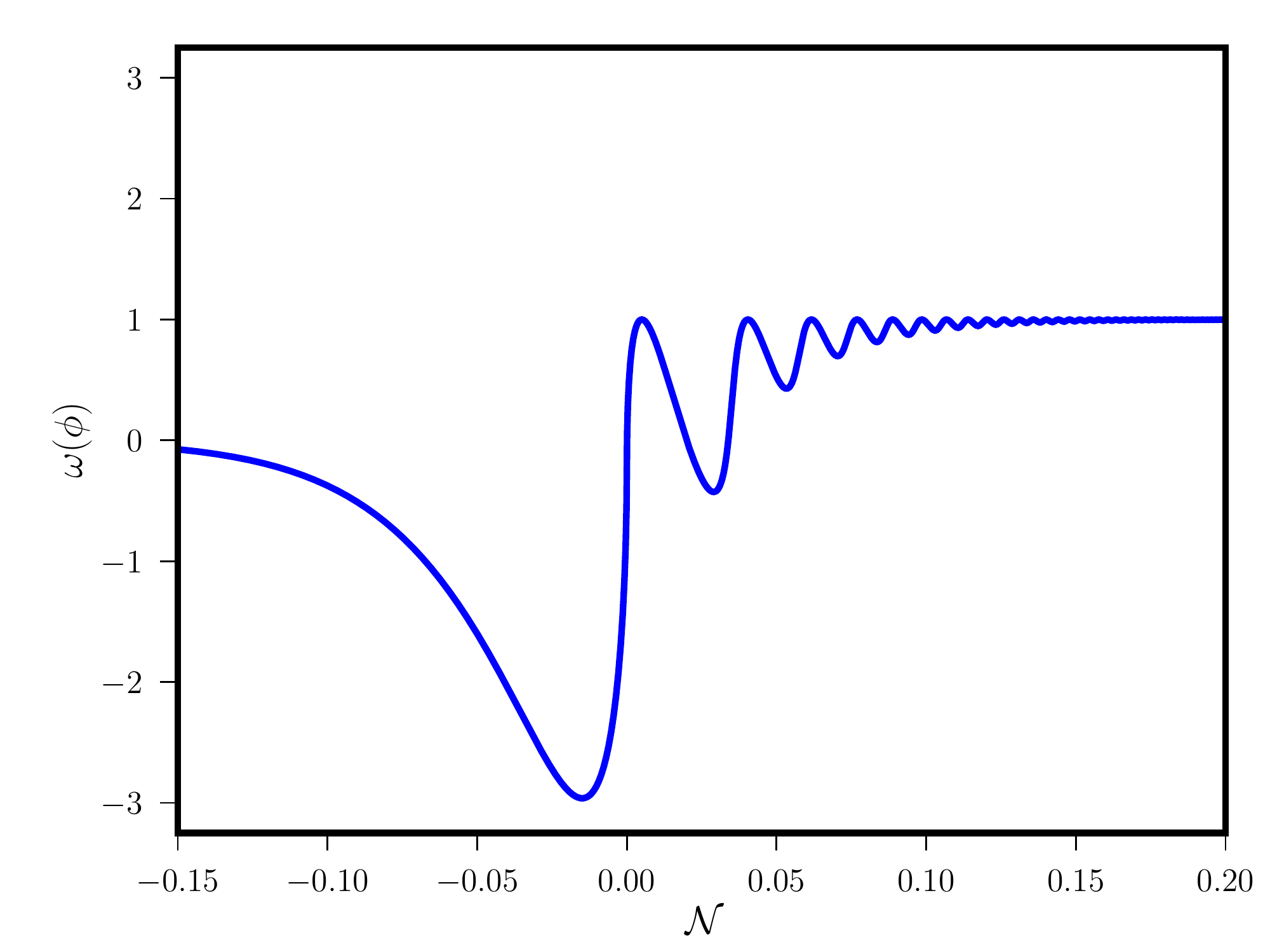}
	\caption{The slow-roll parameter $\epsilon_b$ (top-left), Hubble parameter $H_b$ (top-right), the coupling function $f(\phi)$ (bottom-left) and $\omega(\phi)$ (bottom-right) in the non-minimal theory \eqref{Eq:NonMinAc} are plotted for the chaotic bouncing model with $\alpha = 2$ with respect to the e-N-fold time convention $\mathcal{N}$ during and after the bounce. One can see that the bouncing solution is {\it asymmetrical} and the time-average of the slow-roll parameter in non-minimal theory during reheating is $3/2$, identical to the minimal theory. Also, the coupling function $f(\phi)$ as well as $\omega(\phi)$ become nearly unity in the reheating epoch.}\label{Fig:Coupling}
\end{figure*}

Let us now focus on the remaining issues: obtaining a non-singular bouncing phase as well as an exit mechanism from bounce to reheating epoch. To understand this, one can look into the relations of Hubble parameters and the slow-roll parameters in the two conformal theories: 

\begin{eqnarray}\label{Eq:ConfRel}
	H_b = f\, H_I \,(1 - \frac{f_{,N}}{f}), \quad	\epsilon_b = \frac{\left(\epsilon_I - \frac{f_{,N}}{f}\right)}{\left(1 - \frac{f_{,N}}{f}\right)} - \frac{\frac{f_{,N}^2}{f^2} - \frac{f_{,NN}}{f}}{\left(1 - \frac{f_{,N}}{f}\right)^2},\quad
\end{eqnarray}

\noindent  with $\epsilon_I \equiv -H_{I, N}/H_I,~\epsilon_b\equiv-H_{b, \mathcal{N}}/(\mathcal{N} H_b)$. $N$ is the e-fold time convention in the minimal inflation theory: $a_I(N) \propto e^N$, whereas, $\mathcal{N}$ is the e-N-fold time convention, which is defined as $a_b(\mathcal{N}) \propto \exp \left(\mathcal{N}^2/2\right)$. The subscripts $N$ and $\mathcal{N}$ denote the derivatives with respect to themselves. When the field $\phi$ is high and far away from the minima of the potential $V_I(\phi)$, it slowly rolls down the potential towards minima with the solution $\phi_{, N} \simeq - \Mpl^2\, V_{I, \phi}/V_I $. This immediately implies $(1 - f_{,N}/f) \simeq -\alpha$ with $\epsilon_b \simeq (\alpha + 1)/\alpha$ which corresponds to $H_b<0$ for $\alpha > 0$. During this phase, the coupling function is extremely small but positive, while $\omega(\phi)$ is extremely small and negative. This phase is represented by the solutions \eqref{Eq:scaleb} and \eqref{Eq:coupling} and it ensures that the Universe in the non-minimal theory is contracting with the scale factor solution \eqref{Eq:scaleb}. In this epoch, the large scales of cosmological interests leave the Hubble horizon. However, when $\phi$ approaches close to the minima of the potential $V_I(\phi)$, the solution of the scale factor \eqref{Eq:scaleb} does not hold and $(1 - f_{,N}/f)$ starts increasing, and at the minima of the potential, it becomes positive unity. Therefore, before the field reaches to the minimum of the potential, at one point, it crosses the value zero (when $(\alpha + 1)\,\phi_{,N} = - \Mpl^2 \,V_{I, \phi}/V_I$) where $H_b$ vanishes and $\epsilon_b$ diverges to negative infinity. This point corresponds to the required \emph{non-singular bouncing} point where the null-energy condition is violated without the ghost instability, which we achieve automatically by choosing the coupling function as Eq. \eqref{Eq:coupling}. This occurs due to the past asymptotic behavior of the coupling function $f(\phi)$ as well as its non-trivial evolution.

\par Shortly after the bounce, as field $\phi$ reaches the minima, it starts oscillating around it. At this stage, the scalar field couples with other field(s) and decays. Since, at the minima of the potential, the coupling function $f(\phi)$ (as well as $\omega(\phi)$) becomes unity, the non-minimal action \eqref{Eq:NonMinAc} as well as its solution also merges with the minimal action \eqref{Eq:MinAc} and its solution. Therefore, as the Universe in the minimal theory undergoes through the reheating epoch, the non-minimal bouncing theory also experiences similar phenomena, and the difference between these two conformal theories eventually vanishes. \emph{These results suggest that the constructed bouncing model is asymmetric, free from the difficulties mentioned earlier and at the same time, it can be in agreement with the observations.}

Few things to note: first, the onset of the bouncing (contracting) phase is not negatively infinite but determined by the finite onset of the corresponding inflationary phase, i.e., if the inflationary era starts at $\eta_*$ (corresponding field value $\phi_*$), then the contracting phase begins at the same time. This may lead to extremely small value of the coupling function which may appear to be strong coupling, however, since the model is conformal to minimal theory, the above problem do not arise in the non-minimal theory as well. Second, while the inflationary phase corresponds to the contracting phase of the bouncing scenario, non-singular bouncing phase occurs close to the minima of the potential $V_I(\phi)$. Therefore, any well-behaved inflationary model with a graceful exit to reheating phase can lead to such bouncing phenomena.

%ssimilar to the minimal theory, also in the non-minimal theory, reheating occurs. This occurs shortly after the bounce. In this time, along with $f(\phi)$, $\omega(\phi)$ also approaches (oscillatory) to unity and therefore, $H_b$ as well as $\epsilon_b$ approach to $H_I$ and $\epsilon_I$, respectively. This phase is similar to the conventional reheating mechanism and the non-minimal action merges with the standard minimal theory.

%\begin{figure}[!b]
%	\centering
%	\includegraphics[width=.8\linewidth]{eos2.pdf}
%	\caption{Slow-roll parameter $\epsilon_b$ is plotted for the chaotic bouncing model with $\alpha = 2$ with respect to the e-N-fold time convention $\mathcal{N}$. As it can be clearly seen, at the bounce, it diverges, as expected. Shortly after the bounce, it starts oscillating (around $1.5$). This behavior is similar to the conventional reheating epoch.}\label{Fig:eos2}
%\end{figure}

\section{Example - chaotic bounce:} Consider the chaotic inflation with the potential $V_I(\phi) = \frac{1}{2} \,m^2 \phi^2$. If we require scale factor solution in the non-minimal theory to be the  matter bounce, \ie $\alpha =2$, the required coupling function becomes $f(\phi) = \exp \left(- \frac{3}{4}\,\frac{\phi^2}{\Mpl^2}\right).$ We numerically solve the non-minimal theory \eqref{Eq:NonMinAc} with e-N-fold time convention $\mathcal{N}$. Note that the bounce occurs at ${\cal N}=0$, with negative and positive values of ${\cal N}$ corresponding to the contracting and the expanding phases, respectively. In Fig. \ref{Fig:eos1}, we plot the slow-roll parameter $\epsilon_b$ as a function of $\phi$ for different initial conditions (different colors) to demonstrate the attractor behavior as well as the required matter contracting solution. As one can see, all curves start at different points, but quickly approach the desired solution with $\epsilon_b \simeq 3/2$ as $\phi$ decreases, confirming that the bouncing model is indeed stable and the Universe is matter contracting. In Figs. \ref{Fig:Coupling}, we plot the slow-roll parameter $\epsilon_b$ (top-left), the Hubble parameter $H_b$ (top-right), the coupling function $f(\phi)$  (bottom-left) and $\omega(\phi)$ (bottom-right) as functions of $\mathcal{N}$ around and after the bouncing point. From these figures, it can clearly be seen that, shortly after the bounce, the scalar field starts oscillating and decays, and the coupling function $f(\phi)$ (as well as $\omega(\phi)$) approaches unity. This is similar to the conventional reheating phase, as mentioned before.

%As can be shown in Fig. \ref{Fig:eos1}, the initial condition force the system with effective slow-roll $\epsilon_b = -1/\mathcal{N}\, H_{b, \mathcal{N}}/H_b$ parameter  to begin with $0.97$. However, it quickly reaches to the desired value of $3/2$ (\ie the matter contraction), which tells us that the solution is an attractor, and therefore evading the BKL instability. During this time, the Hubble parameter $H_b$ is extremely small and negative. The field climbs the potential approaching towards zero.

%In Figs.~\ref{Fig:eos2} and \ref{Fig:Hubble}, one can see the bouncing phase as well as the reheating phase. At the bounce, while the $\epsilon_b$ diverges in the negative direction, the Hubble parameter $H_b$ changes the direction from negative to positive. During this time, the field reaches the maxima of the potential and afterwards, it starts rolling down the potential. Shortly after that, it reaches the minima and starts oscillating around zero which signifies the reheating era. At this epoch, the coupling function $f(\phi)$ (and hence, the $\omega(\phi)$) approaches to unity (see Fig.~\ref{Fig:Coupling}) and the potential $V_b(\phi) \simeq V_I(\phi)$. These results hints at the merging of non-minimal theory into a minimal one.

\begin{figure}[!t]
	\centering
	\includegraphics[width=.9\linewidth]{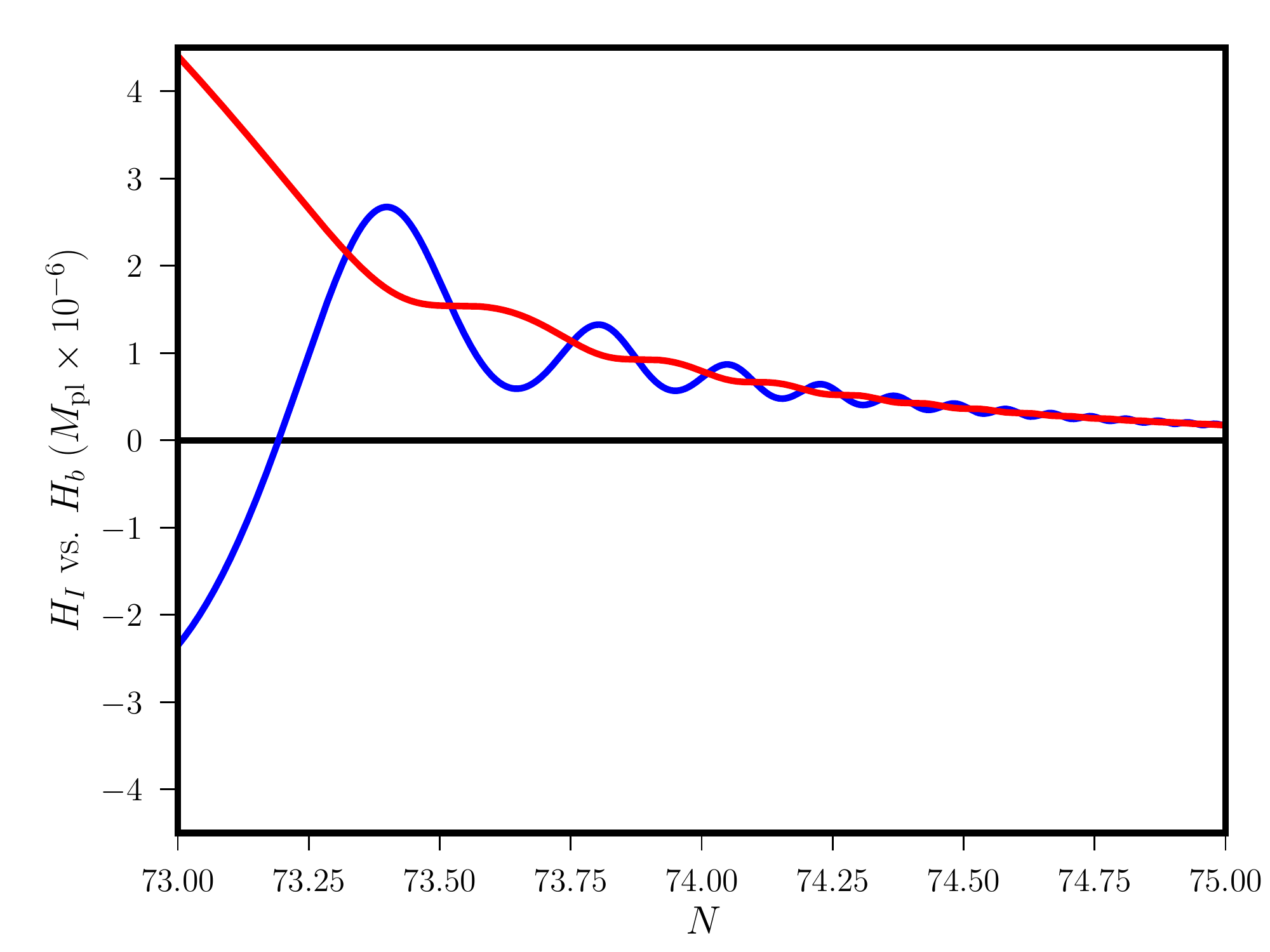}
	\caption{The Hubble parameters $H_I$ in the chaotic inflationary model (red) and $H_b$ in the chaotic bouncing model with $\alpha=2$ (blue) are plotted with respect to the e-fold time convention $N$ before and after the bounce. It is clear that, during reheating, they behave similarly and tend to merge with each other.}\label{Fig:compare_hubble}
\end{figure}

To compare the merging of these two theories into one during and after the reheating phase, one can compare the Hubble parameters in two different theories. By solving the chaotic inflation model with e-fold time convention $N$ and by using \eqref{Eq:ConfTGen}, i.e., ${\rm d}N = \mathcal{N}{\rm d}\mathcal{N} + {\rm d}f/f$, one can easily obtain a one-to-one correspondence in these time parameters $\{N \xleftrightarrow{} \mathcal{N}\}$. Therefore, any parameters can be expressed in any time conventions $N$ or $\mathcal{N}$. In Fig.~\ref{Fig:compare_hubble}, we plot the Hubble parameters in these two theories in e-fold time convention $N$. The blue curve is the Hubble parameter in the non-minimal frame (this is identical to the top-right figure of Fig. \ref{Fig:Coupling}, except the time parameter is now changed to $N$), while red denotes the same in the minimal inflationary theory. $H_b$ vanishes at $N \simeq 73.2$ which represents $\mathcal{N} = 0$, i.e., the non-singular bouncing phase, and as $N$ increases during reheating, the difference between $H_b$ and $H_I$ vanishes. One can also verify that, during reheating, the time-average behavior is same in both frames, i.e., in this case, the effective equation of state is zero (equivalent effective slow-roll parameter  $\langle\epsilon\rangle =3/2$).

However, recent observations ruled out many models including the chaotic inflation, and therefore, the reconstructed chaotic bounce is also not in line with the observations. Despite that, a significant number of inflationary models are still in good agreement with the tighter constraints. For example, the Starobinsky model with $V_I(\phi) = \frac{3}{4}\, m^2\, \Mpl^2 \,\left(1 - e^{-\sqrt{\frac{2}{3}}\, \frac{\phi}{\Mpl}}\right)^2$ is one of them and similar to chaotic bounce, one can also construct a {\it Starobinsky bouncing model} with the coupling function
%By using the relation: $\{N \rightarrow \mathcal{N}\}$ and knowing, during inflation, the earliets  in the case of inflation, it leaves $60$ e-folds before the end of inflation). Since perturbations remain invariant (see \eqref{Eq:pert_rel}), the perturbations in the non-minimal theory remains same as the perturbations in the chaotic inflationary model. This implies that the bouncing non-minimal model satisfies the consistency condition (which has been partly shown in Ref. \cite{Nandi:2019xag}). However, the chaotic inflation model presently is not in agreement with the current observational constraints. However, one can easily obtain the non-minimal bouncing theory from an inflationary model which is well-within the constraint. One such example is the Starobinsky model with $ V_I(\phi) = 3/4\,m^2\,\Mpl^2\,(1 - \exp\left(-2/3\, \phi/\Mpl\right))^2$. In that case, $f(\phi)$ takes the form

\begin{eqnarray}\label{Eq:StarCoup}
	  f(\phi) = \exp \left[ -\frac{3}{4}(\alpha + 1)\, \left(e^{\sqrt{\frac{2}{3}}\, \frac{\phi}{\Mpl}} -  \sqrt{\frac{2}{3}}\, \frac{\phi}{\Mpl} -1\right)\right].
\end{eqnarray}
\noindent  Using the above coupling function, one can easily obtain the non-minimal theory \eqref{Eq:NonMinAc} responsible for the required bouncing solution. In this case, as it is obvious, the reconstructed bouncing model evades all the difficulties including the {\it no-go} theorem.

\section{Conclusions} 
%In order to solve the problems related to constructing a viable bouncing models, we took help of the non-minimal coupling. We, first, countered the difficulty of evading BKL instability as well as the {\it no-go} theorem and found that, a simple non-minimal bouncing model conformal to a viable inflationary background can resolve these two issues, easily. We constructed the model accordingly and discover that, it automatically resolves the issue of having a non-singular bouncing scenario as well. We also noticed that, similar to inflationary dynamics, shortly after the bounce (similar to the end inflation), it naturally oscillates around the minima of the potential which represents the reheating era.  We also found that the during this time, the non-minimal bouncing theory reduces to the minimal one. In short, we show that, for every inflationary model, we can easily construct a bouncing model by using conformal transformation and we can evade the conventional bouncing problems.
While we successfully resolve most of the issues related to the bouncing paradigm by using conformal transformation, we should also stress that, some issues may arise due to the same transformation. For instance, it may not be possible to distinguish between an inflationary model and the corresponding conformal bouncing model \cite{Magnano:1993bd, Francfort:2019ynz, Catena:2006bd, Chiba:2013mha, Saltas:2010ga, Capozziello:2010sc, Kamenshchik:2014waa, Ruf:2017xon, Nandi:2018ooh, Nandi:2019xlj}. However, recently it has also been shown that the reheating dynamics in different conformal theories may behave differently which can help to differentiate the conformal theories \cite{Nandi:2019xve}. Also, the `equivalence' can be broken during and after the reheating epoch as it also depends on how the newly created relativistic particles couple with the gravity, and therefore, the viability of these models may be verified from the future experiments. There are other significant things to note. First, while the value of $\alpha$ we consider as positive, in theory, we can create any form of the scale factor solution using $f(\phi)$. Second, the above procedure holds even for arbitrary field redefinition (e.g., disformal transformation). Third, during reheating, while the average behavior in these conformal theories are the same, one can clearly see the difference in amplitude of the oscillation in the Hubble parameter (and in the slow-roll parameter as well) in Fig. \ref{Fig:compare_hubble}. Fourth, and most importantly, while we provide a class of viable bouncing models associated with inflationary models, we bring a similar problem that plagues the inflationary paradigm: there are many inflationary models, and hence our proposed conformal bouncing models, that remain consistent even with the recent constraints.

To conclude, our novel idea not only leads to solving bouncing theory, but may also open new areas in studying conformally connected theories and its application.

%To conclude, our novel but simple idea leads to the class of first viable bouncing models in the non-minimal frame while the minimal theory is responsible for slow-roll inflation, and it opens a new outlook in looking into conformal frames.   
%We should also mention that, during reheating, as can be seen from Fig.~\ref{Fig:compare_hubble}, despite the average behavior in the minimal and non-minimal theory to be the same, the oscillatory behavior is different (e.g., the amplitude of the field $\phi$), which needs to be studied thoroughly. We reserve the analysis for future work.

\section*{Acknowledgements}
The author thanks L. Sriramkumar, S. Shankaranarayanan, Yi-Fu Cai, and Alexander Vikman for useful discussions and their valuable comments. The author also wishes to thank the Indian Institute of Technology Madras, Chennai, India, for support through the Exploratory Research Project PHY/17-18/874\newline/RFER/LSRI.

%%%%%%%%%%%%%%%%%%%%%%%%%%%%%%%%%%%%%%%%%%%%%%%%%%%%%%%%%%%%%%%%%%%%%%%%%%%%%%%

%\bibliographystyle{elsarticle-harv}
%\bibliography{/home/debottam/Dropbox/Bibliography/Mycollection}

%%%%%%%%%%%%%%%%%%%%%%%%%%%%%%%%%%%%%%%%%%%%%%%%%%%%%%%%%%%%%%%%%%%%%%%%%%%%%%%

\end{document}